# SEMANTIC INFORMATION RETRIEVAL USING ONTOLOGY IN UNIVERSITY DOMAIN

Swathi Rajasurya , Tamizhamudhu Muralidharan , Sandhiya Devi,

Dr.S.Swamynathan

Department of Information and Technology,College of Engineering,Guindy,

Anna University,Chennai-25

rsswathi12@gmail.com,t.amudhu@gmail.com,d27sandhya@yahoo.com

*ABSTRACT:*

Today's conventional search engines hardly do provide the essential content relevant to the user's search query. This is because the context and semantics of the request made by the user is not analyzed to the full extent. So here the need for a semantic web search arises. SWS is upcoming in the area of web search which combines Natural Language Processing and Artificial Intelligence. The objective of the work done here is to design, develop and implement a semantic search engine- SIEU(**Semantic Information Extraction in University Domain**) confined to the university domain. SIEU uses ontology as a knowledge base for the information retrieval process. It is not just a mere keyword search. It is one layer above what Google or any other search engines retrieve by analyzing just the keywords. Here the query is analyzed both syntactically and semantically. The developed system retrieves the web results more relevant to the user query through keyword expansion. The results obtained here will be accurate enough to satisfy the request made by the user. The level of accuracy will be enhanced since the query is analyzed semantically. The system will be of great use to the developers and researchers who work on web. The Google results are re-ranked and optimized for providing the relevant links. For ranking an algorithm has been applied which fetches more apt results for the user query.

## 1.INTRODUCTION:

Conventional web search engines are the most widely used system nowadays for searching and retrieving the results. But the problem is that, the documents and contents are retrieved only based on keywords. This may not provide the most relevant and useful content related to the user query. Here semantics of the query is not considered. It is a mere keyword based search.A user may also require the web services associated with the retrieved content. But the generic search engines do not provide the web services associated with the request automatically. If the query is a location dependent query the apt results relevant to the location may not be retrieved appropriately.

## 1.1 SEMANTIC WEB

The Semantic Web is a Web with a meaning. It describes things in a way that computers can understand. It is an extension to the normal Web and is not about links -relationships between things and its properties. Conventional Web consists of human operator and uses computer systems for tasks like finding, searching and aggregating whereas Semantic Web is the one understood by computers, does the searching, aggregating and combining information without a human operator. It is easily processable by machines, on a global scale. It is the efficient way of representing data on the World Wide Web.

### 1.1.1 LIMITATIONS IN CONVENTIONAL WEB SEARCH

Conventional web search engines are the most widely used system nowadays for searching and retrieving the results. But the problem is that, the documents and contents are retrieved only based on keywords. This may not provide the most relevant and useful content related to the user query. Here semantics of the query is not considered. It is a mere keyword based search.

A user may also require the web services associated with the retrieved content. But the generic search engines do not provide the web services associated with the request automatically. If the query is a location dependent query the apt results relevant to the location may not be retrieved appropriately.

### 1.1.2 NEED FOR SEMANTIC WEB:

The above limitations present in the conventional web search is overcome by building a semantic search engine, thereby analyzing the meaning of the query and providing more appropriate results to the users through keyword expansion.

## 1.2 OBJECTIVE

The aim of the project is to design and implement a semantic search that retrieves the search results analyzing the context and semantics of the query. The semantic search retrieves the most relevant results for the queries under university domain. This search is made possible by construction of a strong ontology which forms the knowledge base. The system eliminates the irrelevant results by forming refined queries and ranking the retrieved links.

## 1.3 PROBLEM DEFINITION

In this information age, it is a deplorable state that despite the overload of information, we regularly fail to locate relevant information. Particularly, in the field of education, several terabytes of content related to various educational institutions such as universities, colleges are uploaded on the internet every week, and the demand for such resources is always on the rise. But access to this information using a generic search engine is not satisfactory in terms of the relevance of links and the overtime on bad links. This can be attributed to several factors, the most important being the absence of identification of context and semantics of the user query in fetching the required results.

In order to overcome these critical issues the proposed system **S**emantic **I**nformation **E**xtraction in **U**niversity **D**omain(**SIEU)** is designed. SIEU retrieves the semantically relevant results for the user query by considering the semantics and context of the query. The Semantics of the query is analyzed by means of the following procedures:

- The user query is initially analyzed grammatically and syntactically by parsing.
- The related synsets for the keywords in the query are retrieved.
- The domain related keywords in the ontology are retrieved to form the refined query.

The results obtained in SIEU are more relevant by adopting the following procedure

- The refined queries that serve as the input for the search engine are formed based on the semantic analysis of the user query.
- The web links retrieved for all the newly formed refined queries are re-ranked based on the domain specific information.

In this way SIEU provides a semantic search that retrieves the appropriate results for the user query.

## 1.4 SCOPE

- Provides an exclusive search service for university related information on the web
- People belonging to different domain can retrieve university related information in an easier way.

- The web results are ranked and more appropriate to the user query.
- The users of this system are provided with the satisfactory results.

This system can also be implemented in a mobile device

## 2. LITERATURE SURVEY

. Semantic Web Searches are an upcoming trend in WWW search. They let knowledge workers concert their efforts and provide a high degree of relevancy and accuracy. Semantic information extraction can be achieved through a multitude of approaches. In this section, we present a survey of some of the existing systems and highlight their unique features.

### 2.1 EXISTING SYSTEMS

Semantic Information Retrieval has become the core part of any search engine. Many papers deal with SWS that uses the OWL language for constructing ontology. DySE System (Dynamic Semantic Engine) **[1]** implements a context-driven approach in which the keywords are processed in the context of the information in which they are retrieved, in order to solve semantic ambiguity and to give a more accurate retrieval based on user interests. DySE splits the user query into subject keywords and the domain specific keywords. It uses a dynamic system that constructs ontology dynamically and uses that as a knowledge base. The DSN (Dynamic Semantic Network) is created by the *DSN Builder*, which generates it from Word Net by means of the *domain keyword* submitted by the user during query submission. In this way a relevance assessment is made in order to compare the results. Ontology Construction in Education Domain **[2]** deals with the construction of Ontology for specific University constructing instances specifically. Here the usage of Protégé tool for constructing the ontology is illustrated. It states the various issues which play a key role in realizing the vision of semantic web such as XML (Extensible Markup Language) and XML Schema, RDF(Resource Description Framework and RDF Schema, URI(Uniform Resource Identifier), Unicode and SPARQL(Standard Protocol for RDF Query language), Search Engines and Agents, and Ontology etc. Ontology development is the objective of the above system and it has provided the guidelines to work in it with education domain as example. Query sentences as semantic networks **[3]** paper describes procedure for representing the queries in natural language as semantic networks. Here a syntactic analysis of the query is done by parsing the query using Stanford parser to tag each and every word with their corresponding parts of speech. Candidate set generation is an important method used here. A plain text based and word net based comparisons are done to match the related concepts in the ontology. There are several ways to information from ontology. Semantic Information Retrieval System **[4]** is mainly concerned with retrieving information from a sports ontology using the **SPARQL** query language. Here specific information is retrieved from the ontology. The sports related information is queried from the ontology and it is done using SPARQL language. It provides a basement for any further research to achieve intelligent fuzzy retrieval of sport information through fuzzy ontology. The pages retrieved from web search needs to be ranked for getting more relevant links. A Relation-Based Page Rank Algorithm for Semantic Web Search Engines **[5]** proves that relations among concepts embedded into semantic annotations can be effectively exploited to define a ranking strategy for Semantic Web search engines. This sort of ranking behaves at an inner level (that is, it exploits more precise information that can be made available within a Web page) and can be used in conjunction with other established ranking strategies to further improve the accuracy of query results. With respect to other ranking strategies for the Semantic Web, this approach only relies on the knowledge of the user query, the web pages to be ranked, and the underlying ontology. Thus, it allows one to effectively manage the search space and to reduce the complexity associated with the ranking task.

The overview of the existing systems gives multitude approaches for semantic information extraction. Though these above systems perform a semantic analysis, it has been implemented in a more generic way. Hence in order to further enrich this process to retrieve more promising results a system has been proposed for queries relating to university domain (SIEU).In this proposed system, in combination with some of the above said methodologies, some more procedures have also been added to perform semantic information extraction in a better way.

## 2.2 CONTRIBUTIONS

SIEU has been designed for retrieving promising results for the queries under university domain. Here after performing a syntactic and semantic analysis of the user query, with the keywords extracted from ontology the refined queries are formed and sent to the search engine. Once the web links are obtained after sending it to a conventional search engine, a re-ranking algorithm has been proposed which justifies that the most relevant web links are filtered and ranked with higher importance and then the less relevant links are produced. Our system also classifies that if it is a location based user query, an analysis is done based on certain keywords in the input query and is separately processed for fetching the most apt results for what the user has asked.

## 3. INITIAL PROCEEDINGS

Ontology is "a formal explicit specification of a shared conceptualization". Ontology provides a common understanding of a term and also its relationship with other terms. Thus a hierarchy can be formed with the related terms. Thus considering our domain, each University will express their purposes and functionalities in different terms. The user query should be parsed so that the stop words that are not needed can be removed from the query. By parsing the query the nouns, verbs and other parts of speech can be used separately if needed. Also Word Net can be used to get synsets of the verbs so that the query can be refined more. The nouns can be used to get their related terms and properties from the constructed ontology.

## 4. ARCHITECTURE

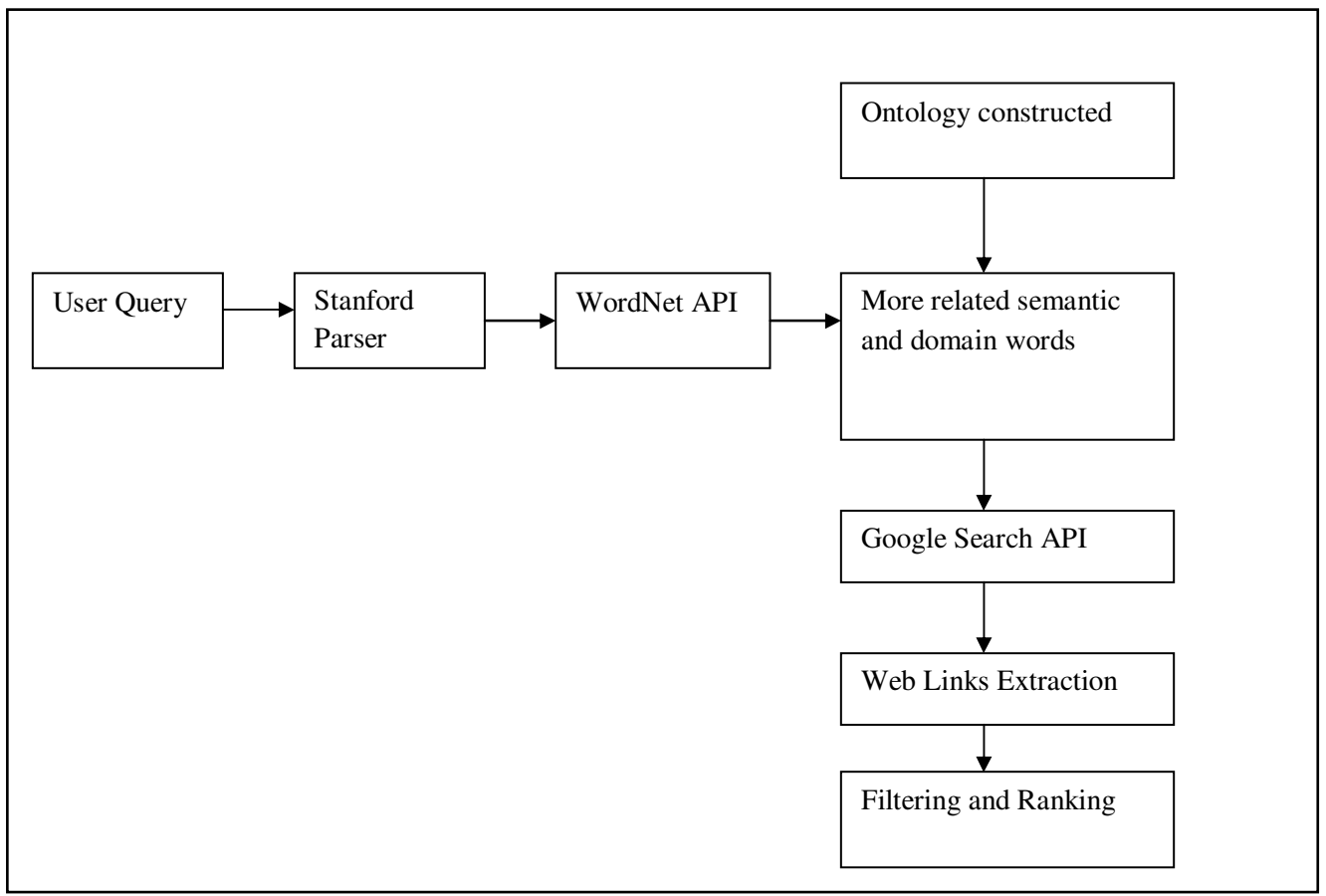

Figure1. Low level design of SIEU

**4.1 HIGH LEVEL DESIGN**

The three major components of the SIEU System are as follows-
- Ontology Construction
- Refined query formation
- Ranking of retrieved links

### 4.1.1 Ontology Construction

This component describes the construction of ontology which forms the knowledge base of SIEU system. The concepts related to university domain are gathered from various university web sites and from various other sources such as Word Net. These concepts are clustered in a hierarchical form in ontology which serves as database for the keywords related to university domain. These keywords are used for the formation of the refined query.

### 4.1.2 Refined Query Formation

The query given by the user is refined to provide a better search result by means of this component. In this component the query given by the user is parsed to identify the parts of speech of the words contained in the query. Then the related synsets for keywords contained in the query are retrieved. The domain keywords that are semantically related to the query are extracted from the ontology built. This step results in the retrieval of more number of semantically related words. These domain keywords are then used for the formation of refined queries. These refined queries are queries with expanded keywords and that has more semantic relevance involved.

**Modules**:
- Query parsing
- Synsets retrieval
- Keywords extraction
- Refined query formation

### 4.1.3 Ranking of retrieved links

This component produces a set of web links that are semantically related to the user query. The set of refined queries serves as an input for the web search which produces the web links corresponding to the given queries. These retrieved web links are filtered and then ranked. The web links are ranked based on the semantic relevance which is attained by means of the extracted domain keywords from the ontology.

**Modules**:

- Retrieval of web links
- Ranking of retrieved links

## 5. DETAILED DESCRIPTION

### 5.1 ONTOLOGY CONSTRUCTION: Knowledge Base

ONTOLOGY, a formal representation of knowledge as a set of concepts within a domain forms the knowledge base for our project that is constructed based on the concepts related to the university domain. By referring through various university websites a handful of information is gathered and based on that a strong ontology is constructed taking into consideration, the various important areas under university domain.

TOOL USED: Protégé 4.1

### 5.2 USER INPUT:

The user of the system enters a query related to university domain in natural language. The expected output of this query is the semantically relevant web links. The irrelevant links are filtered out.

### 5.3 PARSING OF INPUT QUERY:

The input query given by the user is initially parsed by means of the parser. The parsing is done to analyze the query syntactically which determines the part of speech of each and every word in the query. In this way the given query is analyzed grammatically.

TOOL USED: Stanford Parser

### 5.4 WORDNET:

The output obtained from the parser is sent to the wordnet to get the related synsets of various words contained in the query. So here semantically related words are obtained from the output of the wordnet.

TOOL USED: Wordnet API

## 5.5 EXTRACTION FROM ONTOLOGY:

This process involves of more importance where the information related to the given user query is extracted from the built ontology. The initially given query after passing through the Stanford parser and wordnet a set of classified and semantically analyzed words are obtained. These words are matched with the concepts contained in the ontology to get a set of more related key words. At the end of this process we get a collection of words which are semantically related and domain specific key words.

TOOL USED: Jena API

## 5.6 FORMATION OF REFINED QUERY:

Next process involves of query formation with these collection of words. Permutations and combinations are needed to form various refined queries from the words obtained. The queries formed will be more refined and will fetch more semantically related web links on passing these queries as input to the search engines. The refined queries are sent to search API which fetches the web links related to the user query.

TOOL USED: Google Search API

## 5.7 RANKING OF WEB PAGES:

The web links obtained after passing the refined queries to the search API are now filtered and ranked to make it more refined. If any of the web links are not relevant to the given query they are filtered out. Ranking is applied to all web links obtained from all possible queries formed under permutation. On applying ranking the web links are re-ranked in the appropriate order of semantic relatedness.

TOOL USED: Ranking Algorithm

# 6. RESULTS OBTAINED IN EVERY STAGE:
## 6.1 SAMPLE QUERY ENTERED BY THE USER:
The user enters the query in the interface provided by our system.

**SIEU**

Semantic Information Extraction in University Domain

[search box: "list the teaching staff in anna university"] [search]

### 6.2 PARSING OF THE INPUT QUERY:

The query given by the user is parsed by means of Stanford parser and the output is:

```
list the teaching staff in anna university parsing NN list DT the NN teaching NN staff IN in NN anna NN university
chunking NP list NP the teaching staff NP anna university
```

### 6.3 RETRIEVAL OF SYNSETS FROM WORDNET:

Now the related synsets for the words present in the query are retrieved from the wordnet.

```
The following synsets contain 'provide' or a possible base form of that text: supply
The following synsets contain 'doing' or a possible base form of that text: make
```

### 6.4 EXTRACTION OF DOMAIN KEYWORDS FROM ONTOLOGY:

The domain keywords that are semantically related to the words in the query are extracted from ontology.

```
faculty
staff
employee
people
list
teaching
anna
university
```

## 6.5 WEB LINKS RETRIEVED:

### 6.5.1 User Query:

> Provide the Faculties in Computer Science Department Anna University.
>
> About 270,000 results (0.14 seconds)
>
> **Department of Computer Science and Engineering - Anna University** - 1:20am
> 13 Feb 2010 ... The **Department** functions under the **Faculty** of Information and ... The **Department provides** state of the art **computing** facilities to the ...
> cs.annauniv.edu/ - Cached - Similar
>
> [PDF] **Brochure - ANNA UNIVERSITY, CHENNAI 600 025**
> File Format: PDF/Adobe Acrobat - Quick View
> This program **provides** opportunities for **faculty** members currently ...
> www.annauniv.edu/qip/brochure.pdf - Similar
>
> **Anna University - Department of Mathematics**
> The **Department** of Mathematics is one of the largest units of **Anna University** ...
> www.annauniv.edu/Maths/index.html - Cached - Similar
>
> ➕ Show more results from annauniv.edu
>
> **Department Of Computer Science Anna University** - Quote Info
> **Department Of Computer Science Anna University** in US. Boasting 18 **faculty** members and set on a major growth trajectory, the **Department** has strong research ...
> ffr.dnsdot.net/department-of-science-university-76308.html - Cached

### 6.5.2 With Refined Query:

> Provide the people or Faculty in Computer Science Department Anna University.
>
> About 299,000 results (0.15 seconds)
>
> ▶ **Department of Computer Science and Engineering - Anna University** - 3:38am
> 20 May 2010 ... **anna university, computer science** and engineering, **department** of **computer science** and engineering, DCSE, **Anna University**, Anna, AU.
> cs.annauniv.edu/People/professor.html - Cached - Similar
>
> **Department of Computer Science and Engineering - Anna University** - 1:20am
> The **Department** functions under the **Faculty** of Information and ...
> cs.annauniv.edu/ - Cached - Similar
>
> ➕ Show more results from annauniv.edu
>
> **People | Department of Computer Science | University of Pittsburgh**
> **People** are what make Pitt's **Computer Science Department** unique. ... She decided to leave her **faculty** position in the **Department** of **Computer Science** at the University of ... degree in Electrical Engineering from **Anna University**, India. ...
> www.cs.pitt.edu/people/index.php - Cached - Similar
>
> **Computer Science Department - Faculty**
> I managed a team of 25 **people** writing SCM application on AS/400 and ASNA . ... I am working as Assistant Professor in **department** of Computer Science & Engg. ..... I have completed M.E. (Software Engineering) from **Anna University** Chennai ..... I feel that DCE will **provide** me with a platform from where I can not only ...
> www.gurgaon.dronacharya.info/CSEDept/facultyprofile.html - Cached

# 7. PERFORMANCE STUDIES

## 7.1 MEASURES USED

Recall:

Measure of how much relevant information the system has extracted (coverage of system).

$$RECALL = \frac{\text{\# of relevant links given by the system}}{\text{Total \# of relevant links in GOOGLE and SIEU}}$$

The Recall calculated here is the relative recall in which performance is compared in relative to the google search engine.

Precision:

Measure of how much of the information the system returns is correct (accuracy).

$$PRECISION = \frac{\text{\# of relevant links given by the system}}{\text{Total \# of links retrieved}}$$

## 7.2 SIEU VS GOOGLE

When our proposed system SIEU was tested, a marked improvement in performance was observed for most of the queries as a result of semantic analysis, although a small fraction of them had negative and similar performance with a generic search engine.

The table 5.11 shows the samples of the performance levels of our system comparing it with Google.

Table 1. Precision and Recall values of sample queries

| SAMPLE QUERIES | GOOGLE | | SIEU | |
|---|---|---|---|---|
| | PRECISION | RECALL | PRECISION | RECALL |
| colleges for doing M.B.A | 0.68 | 0.44 | 0.87 | 0.5 |
| teaching staff in computer science department in Anna university | 0.62 | 0.41 | 0.86 | 0.6 |
| professors with more number of publications in IIT in department IT | 0.68 | 0.5 | 0.78 | 0.54 |
| last date to apply for M.S in Stanford university | 0.56 | 0.43 | 0.77 | 0.57 |
| financial aid offered for summer internships in UK | 0.75 | 0.46 | 0.87 | 0.56 |
| Deadline for payment of fees for M.B.A course in sastra university | 0.53 | 0.31 | 0.77 | 0.56 |
| Associations formed for students in California university | 0.7 | 0.45 | 0.88 | 0.55 |
| Provide me the details of the chairman of board of committee members | 0.66 | 0.52 | 0.73 | 0.6 |
| Research areas in IIT where foreign collaborations exists | 0.56 | 0.5 | 0.68 | 0.54 |
| Details about the facilities available in research institutions of delhi university | 0.6 | 0.55 | 0.57 | 0.55 |
| Provide me the information about the correspondence students of MIT | 0.7 | 0.45 | 0.7 | 0.56 |
| Road maps to visit the campus of Stanford university | 0.65 | 0.55 | 0.78 | 0.61 |
| Information regarding the universities in abroad which provides internship in accounting | 0.68 | 0.46 | 0.60 | 0.45 |
| Procedure to apply online for M.S in U.S university | 0.74 | 0.45 | 0.79 | 0.65 |
| How far is tagore university located | 0.76 | 0.58 | 0.81 | 0.59 |
| from anna nagar | | | | |
| What are colleges located near by tambaram for doing regular M.E course | 0.67 | 0.45 | 0.83 | 0.55 |

The above table depicts that the precision value of our system SIEU is higher than the values obtained in google search engines. The relative recall values estimates the retrieval effectiveness between Google and our system. The more relative recall values of out system shows that SIEU is more effective in retrieval than Google search engine.

**7.3 Precision Recall curve**

Figure 5.1 shows the precision Vs recall graph for SIEU and GOOGLE system .The graph is drawn taking the first 5 queries into consideration, their corresponding precision and recall values are plotted for both Google and SIEU systems. The precision recall curve in the graph clearly depicts that SIEU system retrieves the accurate links for the user query based on semantic relatedness. The values of precision and recall depicts the performance of SIEU system.

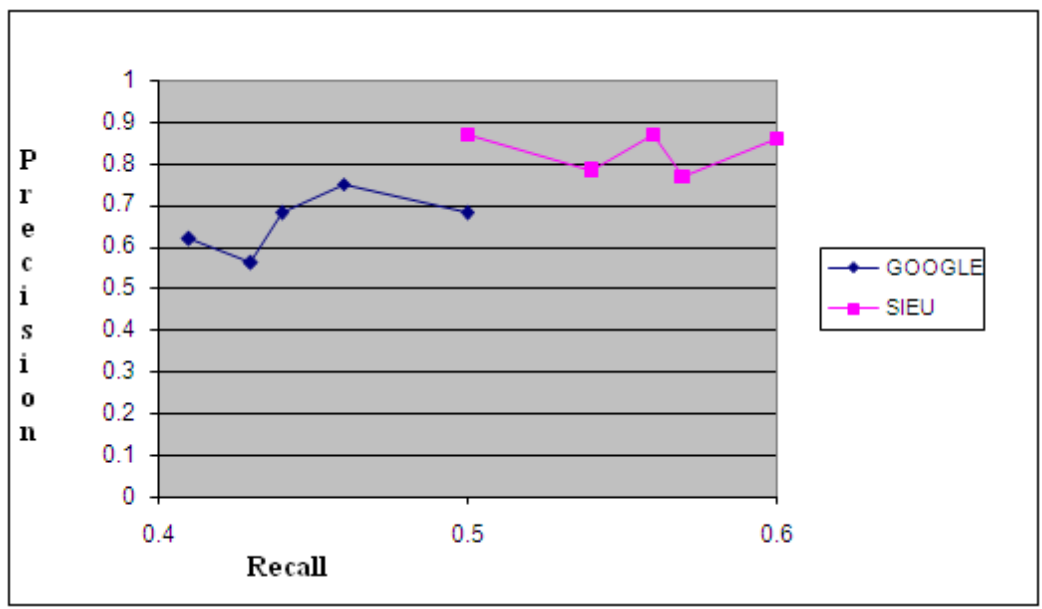

**Figure 1. Precision Vs Recall graph for SIEU Vs GOOGLE**

**Table 2. Average precision and recall of SIEU**

|  | SIEU | GOOGLE |
|---|---|---|
| **AVERAGE PRECISION** | 0.79 | 0.64 |
| **AVERAGE RECALL** | 0.55 | 0.48 |

Table 5.12 shows the average precision and recall of SIEU and Google. From this we can infer that the higher value of average precision and recall for our proposed system SIEU when compared to Google depicts that our system has a better performance and accuracy in retrieving the results than the generic search engines. The recall values depict the coverage of the system. The average relative recall value of SIEU system is also higher compared to that of the Google. The higher value denotes the best coverage of our system compared to the generic search engines.

## 8. CONCLUSION:

Semantic relevant information has been retrieved as a result of this system. To add on to it many more services are to be added. Location based information retrieval is an additional feature where we have planned to use google maps for this purpose by means of which our system gets enhanced with this location independent feature. And invocation of web services may be provided to the users if incase there exists a web service related to their query which could be made possible with the help of RSS feeds. Thus we believe an information system with these enhanced features will be developed under university domain.

# APPENDIX A

- **Stanford parser**

    Stanford parser is a natural language parser that works out the grammatical structure of sentences, for instance, which groups of words go together (as "phrases") and which words are the subject or object of a verb. Tagging process is done by the parser where the words are tagged by their parts of speech.

- **Word net**

    Word Net is a large lexical database of English. Nouns, verbs, adjectives and adverbs are grouped into sets of cognitive synonyms called as synsets, each expressing a distinct concept.

Java API for Word Net Searching (JAWS) is an API that provides Java applications with the ability to retrieve data from the Word Net database

- **Jena Framework Api**

  Jena is a Java framework for building Semantic Web applications. It provides a programmatic environment for RDF, RDFS and OWL and includes a rule-based inference engine. Jena is open source.

- **Google search Api**

  Google Search Api coordinates a search across a collection of search services. It provides all kinds of search such as local search, web search, News search, Video search etc. Google Api loader loads this search Api which provides the search results from the generic search engines.

- **HTML Parser**

  HTML Parser is a Java library used to parse HTML. Primarily it is used for transformation or extraction. It features filters, visitors, custom tags. It is a fast, robust.It is used to extract the meta tags from the web pages.